# Similarity of the Boson Peaks in Disordered Systems to the van Hove Singularities in Regular Crystals

**I.A. Gospodarev, V.I. Grishaev, O.V. Kotlyar, K.V. Kravchenko, E.V. Manzhelii, E.S. Syrkin, S.B. Feodosyev, I.K. Galetich, A.V. Yeremenko**

B.I. Verkin Institute for Low Temperature Physics and Engineering NAS of Ukraine, 61103, 47 Lenin ave., Kharkov, Ukraine

Phonon spectra of solid substitutional solutions with finite concentrations of impurities were analyzed on the microscopic level. The local phonon densities of impurity atoms were calculated, in particular the formation of quasilocal vibrations and their evolution with increasing of the concentration of impurities were investigated. Modification of the local spectral densities of atoms of the host lattice by impurities and manifestation of the phonon Ioffe-Regel crossover (scattering of fast propagating phonons on quasi-localized vibrations) were analyzed. It is shown that such scattering causes the manifestation of the features such as "boson peak" in the phonon spectrum of solid substitutional solutions. A commonality of the physical nature of such singularities in disordered structures and van Hove singularities in ideal crystals was established, notably, reducing of the group velocity of acoustic phonons due to their scattering on slow phonons.

## 1. Introduction

It is well known that heavy or weakly bound to the atoms of the host lattice impurities enrich the low-frequency region of the phonon spectrum and lead to a significant change in low-temperature thermodynamic and kinetic characteristics (see, for example, [1-7]). In the linear in concentration approximation the changes of the phonon spectrum due to both the heavy and light impurity atoms are well described by the Lifshitz theory of regular perturbation [8-10]. In particular, formation of the so-called quasi-local vibrations (QLV) i.e. the resonance peaks in the low-frequency phonon spectrum by heavy or weakly bound impurities, and their contribution to the low-temperature vibrational characteristics were predicted [11] and analyzed in detail theoretically (see for example, [12,13]) and experimentally (see, e.g., [14-16]). QLV are localized near the impurity atoms and their formation is very similar to the occurrence of discrete vibrational levels (local oscillations) outside the continuous spectral band of the basal lattice in the presence in crystal of light or strongly coupled impurities. However, there is an important fundamental difference between local and quasilocal vibrations, manifested under increasing concentration of impurity atoms. Local vibrations are the poles of the Green function of the perturbed crystal, and their amplitudes decay exponentially with distance from the impurity. Being located outside of the band of quasi-continuous spectrum, these vibrations do not interact with the phonon modes of the host lattice. With an increasing content of either light or strongly coupled impurities their effect upon phonon spectrum can be defined as an expansion in the concentration (see, for example, [17]). Thus, an increasing content of light impurities leads to appearance of sharp resonant peaks in phonon spectrum at frequencies coincident with those of local vibrations which are attributed to the vibrations of the isolated impurity atom pairs and eventually, regular triangles and tetrahedra [18,19]. QLV are not the poles of Green function but common maxima in phonon density of states which do not violate its analyticity. Although, as it is shown in the next section, these peaks are formed just by vibrations of impurity atoms, QLV interact with the phonon modes of the host lattice. Therefore, even at low enough concentrations (about few percent) of either heavy or weakly coupled impurity atoms the significant modification of the entire phonon spectrum occurs. This transformation of the entire phonon spectrum can not be described by the expansion in the concentration of impurities. As an example of the influence of the described previously transformation of the phonon spectrum upon thermodynamic quantities we refer the "two-extreme" behaviour of the temperature dependence of the relative change in the low-temperature heat capacity of Kr-Ar solution [20, 21] (in krypton matrix interaction between argon impurities becomes weaker). This temperature dependence of heat capacity is unexplained by a superposition of contributions of isolated impurities, impurity pairs, triples and etc., without taking into account the restructuring of the entire spectrum [22]. The restructuring of the phonon spectrum of the crystal and delocalization of QLV at finite impurity concentrations were considered in [23-26] within the coherent potential approximation.

QLV usually occur in the frequency range where corresponding wavelengths of acoustic phonons of the host lattice become comparable with the average distance between the defects (the so-called disorder parameter) even at low concentrations of impurity atoms. Therefore an interaction of QLV with rapidly propagating acoustic phonons of the host lattice (so-called *propagons*) appears as the Ioffe-Regel crossover, as it is shown in [27-29], and can lead to the formation of boson peak (BP). BP is the anomalous override of low-frequency phonon density over Debye density [30-32]. BP were revealed in the Raman and Brillouin scattering spectra [33,34] as well as in experiments on inelastic neutron scattering [35] by peaks in the frequency dependence of ratio of the phonon density of states $\nu(\omega)$ or the scattering intensity $I(\omega)$ to the square of the frequency $\omega$. These peaks appear in the low-frequency region of vibrational density of states in the frequency range between 0.5 and 2 THz [36, 37], i.e. far below the Debye frequency, while at the BP frequency a transition occurs from fast-propagated low-frequency phonons (propagons) with dispersion law close to the acoustic one, to the so-called *diffuzons*, i.e. phonons, whose propagation is hampered by scattering on localized states appearing at higher frequencies [38]. The similarity of the boson peak in disordered systems (e.g. glasses and substitution solid solutions) with the first van Hove singularity in the crystal structures is noted in [39,40]. BP were also observed in polymeric and metallic glasses [41- 46].

However, the conditions of QLV formation and, in particular, their behaviour at a finite concentration of impurity atoms and transformation with its growth were not yet sequentially analyzed at the microscopic level. But just such an analysis allows explaining the nature of BP and provides general description for all features of low-frequency phonon spectra of real crystals and solid solutions.

The second section is devoted to analysis of the formation conditions and characteristics of QLV at the microscopic level. The evolution of QLV with increasing concentration of impurity atoms has been investigated. Within a framework of realistic crystal lattice low-frequency anomalies of the phonon spectrum, caused by both the vibrations localized on the impurities and by their scattering of the fast acoustic phonons due to vibrations of host atoms, were calculated and analyzed.

In the third section it is demonstrated, that BP originate from additional dispersion of fast acoustic phonons due to their scattering on QLV, and the similarity of BP in disordered systems with van Hove singularities in regular crystal systems is analyzed.

## 2. Quasilocal vibrations. The formation and evolution with increasing concentration of impurity atoms

For crystals with simple and highly symmetrical lattice the dispersion law of low-frequency phonons, i.e. phonons with frequencies lying well below the first van Hove singularity, is close to the sound dispersion law $\omega \approx sk$ ($\omega$ is frequency, $k$ is a module of the wave vector $\mathbf{k}$ and $s$ is sound velocity) and vibrational characteristics of ideal systems are rather reasonable described within the Debye approximation. In the low-frequency region, the phonon density of states is the same as the Debye one, i.e. at $\omega \to 0$ the function $\nu(\omega)$ falls down as $\omega^2$. Therefore, the introduction of various defects into crystal can significantly enrich low-frequency region of the phonon spectrum and lead not only to quantitative but also qualitative changes in the behaviour of low-temperature vibrational characteristics.

At low impurity atoms concentrations $p \ll 1$, vibrational characteristics of the solid solution can be described within the linear in $p$ approximation:

$$\tilde{\nu}(\omega) = \nu(\omega) + p\sum_i \Delta\rho_i(\omega) \qquad (1)$$

The summation is performed over all cyclic subspaces with non-zero operator $\hat{\Lambda}$ which describes perturbation of the lattice vibrations by either isolated heavy impurity or weakly coupled one, $\Delta\rho^{(i)}(\omega)$ is the spectral density change in each of these subspaces, $\tilde{\nu}(\omega)$ and $\nu(\omega)$ are the phonon densities of states of solid solution and ideal crystal, respectively. If in each of the cyclic subspaces operator $\hat{\Lambda}$ induces a regular degenerate operator, then the value $\Delta\rho^{(i)}(\omega)$ can be calculated using the spectral shift function [9, 10]. Using the expressions obtained for this function in the J-matrix method [47-51], we obtain:

$$\Delta\rho(\omega) = -\frac{d\xi(\omega)}{d\omega} = \frac{\rho^2(\omega)}{\pi^2\rho^2(\omega) + \left[S(\omega) - \operatorname{Re}G(\omega)\right]} \cdot \frac{d}{d\omega}\left[\frac{S(\omega) - \operatorname{Re}G(\omega)}{\rho(\omega)}\right], \qquad (2)$$

where $\xi(\omega)$ is the spectral shift function [10]; the function $S(\omega)$ describes perturbation by defect and depends on the defect parameters, $G(\omega)$ is the local Green function of an ideal crystal.

If in any cyclic subspace the solution of equation

$$S(\omega) - \operatorname{Re}G(\omega) = 0 \qquad (3)$$

is $\omega = \omega_k$, then in this point vicinity the expression (2) has a resonant character.

$$-\frac{d\xi}{d\omega} = \frac{2}{\pi}\frac{\Gamma}{4(\omega - \omega_k)^2 + \Gamma^2}; \qquad \Gamma \equiv \frac{\pi\rho(\omega)}{\frac{d}{d\omega}\left[S(\omega) - \operatorname{Re}G(\omega)\right]_{\omega=\omega_k}}. \qquad (4)$$

Equation (3) formally coincides with the Lifshitz equation which determines (for naturally different $S(\omega)$ function), meanings of the frequencies of discrete vibrational levels, lying outside the band of quasi-continuous spectrum of the crystal [9, 10]. However, these discrete levels in contrast to the values $\omega_k$ are the poles of the perturbed local Green function. Green function can not have poles within the band of quasi-continuous spectrum. The possibility to determine the QLV frequencies using equation (3) is due to the fact that at low frequencies $|\operatorname{Re}G(\omega)| \gg \operatorname{Im}G(\omega)$ though at $\omega = \omega_k$ the spectral density of an ideal crystal can not be considered negligible for a lot of realistic defect-parameter values.

Let us analyze the quasi-local oscillations due to substitution impurity in FCC crystal with central nearest-neighbour interaction. Interaction of impurity with the host lattice is also considered as purely central and, therefore, as it is noted in the previous section, the perturbation caused by such impurity should be regular and degenerate. Let us consider two cases: the isotopic impurity with mass $m$ four times higher than that of the host lattice (i.e. the mass defect is $\varepsilon \equiv \frac{\Delta m}{m} = 3$) and the impurity atom with a mass equal to the mass of an atom of the host lattice $\varepsilon = 0$, though coupled to the host lattice atoms four times weaker than between each other ($\upsilon \equiv \frac{\Delta\alpha}{\alpha} \equiv -\frac{3}{4}$ is the coupling defect, $\alpha$ is the nearest neighbour interaction force constant). In the first case, operator $\hat{\Lambda}$ induces a non-zero operator only in the cyclic subspace $H^{(\tau_-^5)} \in H$ ($H$ is the space of atomic displacements) which is generated by displacement of the just impurity atom.

The vectors pertained to this subspace transform according to irreducible representation $\tau_-^5$ of symmetry group of the lattice $O_h$ (the notation of [52]). In the given subspace the spectral density of an ideal lattice coincides with its density of states. For an isotopic impurity the function $S(\omega)$ [47, 48] reads:

$$S_{is}(\omega) = -\frac{2}{\omega\varepsilon} . \qquad (5)$$

In the second case, except the subspace $H^{(\tau_-^5)}$ where the function $S(\omega)$ reads:

$$S_w^{(\tau_-^5)}(\omega) = \frac{2}{\omega} + \frac{\omega_m^2(1+\upsilon)}{\omega^3 \upsilon} , \qquad (6)$$

the non zero operators induced by the operator $\hat{\Lambda}$ in cyclic subspaces are those transformed according to irreducible representations, $\tau_+^1$, $\tau_+^3$, $\tau_+^4$ and $\tau_-^4$ of the same group $O_h$ will be different from zero ($\omega_m$ is the upper limit of quasi-continuous spectrum). Over all of these four subspaces $S$ - functions coincide

$$S_w^{(\tau_+^1)}(\omega) = S_w^{(\tau_+^3)}(\omega) = S_w^{(\tau_+^4)}(\omega) = S_w^{(\tau_-^4)}(\omega) = S_w'(\omega) = \frac{16\omega}{\omega_m^2 \upsilon}. \qquad (7)$$

For weakly bound impurity, the function $S_w' \leq S_{lim} = \frac{16\omega}{\omega_m}$, and as shows Fig. 1, equation (3) can has no solutions within the cyclic subspaces $H^{(\tau_+^1)}$, $H^{(\tau_+^3)}$, and $H^{(\tau_+^4)}$. In the subspace $H^{(\tau_-^4)}$, solution is possible for impurity, whose force interaction with atoms of the host lattice is weaker than the interaction of atoms of the host lattice with each other by at least 50 times. Value $\omega_k$ will be virtually the same as the first van Hove singularity (its frequency will be labelled as $\omega^*$).

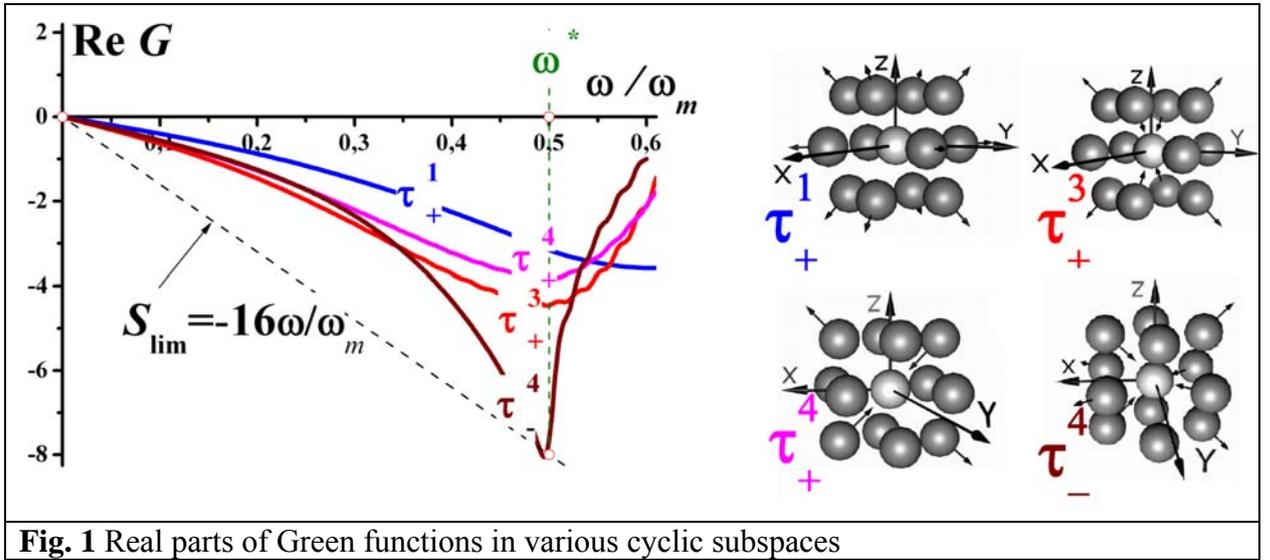

**Fig. 1** Real parts of Green functions in various cyclic subspaces

Therefore, for real values of parameter $\upsilon$ equation (3) has a solution in the subspace $H^{(\tau_-^5)}$ only. This solution for both cases is shown in Fig. 2.

The real part of Green function (curves 2 in both panels) crosses the dashed curves 3, which represent the dependence (5) (the upper panel) and (6) (the bottom panel), in corresponding points $\omega_k$. This figure also shows the spectral densities $\rho^{(\tau_-^5)}(\omega)$ of the ideal crystal, coinciding with its phonon density of states $\nu(\omega)$ (curves 1 (dashed)) and phonon densities of states of the corresponding solid solutions with concentration $p = 5\%$.

The value of the phonon density of states $\nu(\omega_k)$ can not be considered negligible ($\nu(\omega_k) \sim 0.1 \operatorname{Re} G(\omega_k)$). Therefore, as it is seen from Fig. 2, the frequencies of the maxima in the curves 4, though they are close to frequency $\omega_k$ but not coincident with it (especially in the case of weakly bound impurity (bottom panel). For weakly bound impurity one should expect a higher degree of localization of QLV on impurity atoms.

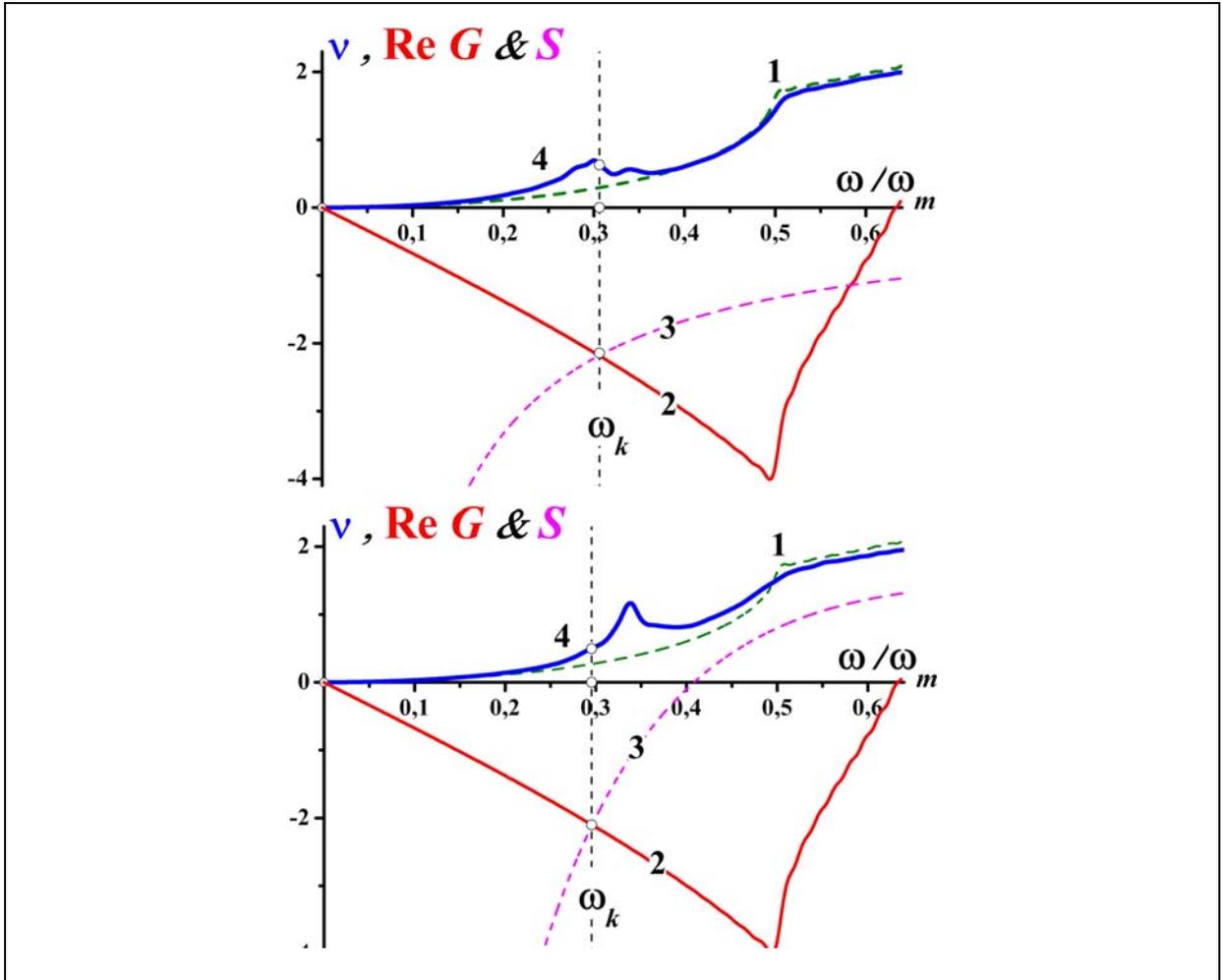

**Fig. 2** Graphical solutions of equation (3) in the cyclic subspace generated by the displacement of the impurity atom: the upper and bottom panels correspond to the heavy isotopic and weakly coupled impurities impurities, respectively.

Modification of low-frequency density of states caused by introduction to the crystal of 5% of heavy (upper panel) and weakly bound (lower panel) impurity atoms ($\Delta\nu(\omega) = \tilde{\nu}(\omega) - \nu(\omega)$ for curve 1) compared in Fig. 3 with the spectral density of an isolated impurity atom (curve 2 is $0.05 \cdot \tilde{\rho}(\omega)$)

$$\tilde{\rho}(\omega) = \frac{2\omega}{\pi} \text{Im}\left(\vec{h}_0^{(\tau_5)}, \left[\omega^2 \hat{I} - \hat{L} - \hat{\Lambda}\right]^{-1} \vec{h}_0^{(\tau_5)}\right), \qquad (8)$$

where $\vec{h}_0^{(\tau_5)} \in H$ is the generating vector of the cyclic subspace $H^{(\tau_5)}$; $\hat{L} + \hat{\Lambda}$ is perturbed operator of lattice vibrations; $\hat{I}$ is the unit operator.

It should be noted that given spectral density has a characteristic resonance Lorentzian form similar to (4), and there are no any singularities of van Hove type on it. The frequencies of peaks on the spectral densities $\tilde{\rho}(\omega)$ (points $\omega_{ql}$) and on the curves $\Delta\nu(\omega)$ coincide with relatively high accuracy, especially for the case of weakly bound impurity. This indicates the strong localization of QLV on impurity atoms. Therefore the frequency $\omega_{ql}$ can be more reasonably than $\omega_k$ determined as the quasi-local frequency. Fig. 3 also represents (curves 3) the spectral correlators of displacements of impurity atoms with their first coordination sphere

$$\rho_{01}^{(\tau_2^5)}(\omega) = \frac{2\omega}{\pi} \mathrm{Im}\left(\vec{h}_1^{(\tau_2^5)}, \left[\omega^2 \hat{I} - \hat{L} - \hat{\Lambda}\right]^{-1} \vec{h}_0^{(\tau_2^5)}\right) = \hat{P}_1(\omega^2) \rho^{(\tau_2^5)}(\omega), \quad (9)$$

where $\hat{P}_1(\omega^2)$ is the polynomial defined by the recurrence relations for the *J*-matrix of the perturbed operator $\hat{L} + \hat{\Lambda}$ [47-48]. Spectral correlator $\rho_{01}(\omega)$ vanishes when $\omega = \omega_E$ where $\omega_E$ is the Einstein frequency of the correspondent subspace ($\omega_E^2 = \int_0^{\omega_m} \omega^2 \nu(\omega) d\omega$). Thus, when $\omega = \omega_E$, a correlation with the first coordination sphere is absent, and the closer $\omega_{ql}$ to the $\omega_E$ frequency, the degree of localization of QLV is greater. As it could be seen from Fig.3, QLV frequency for weakly bound impurity is nearly three times closer to $\omega_E$ than for isotopic one, and quasi-local maximum for a weakly bound impurity has a sharper resonance form than the maximum for heavy isotopic defect.

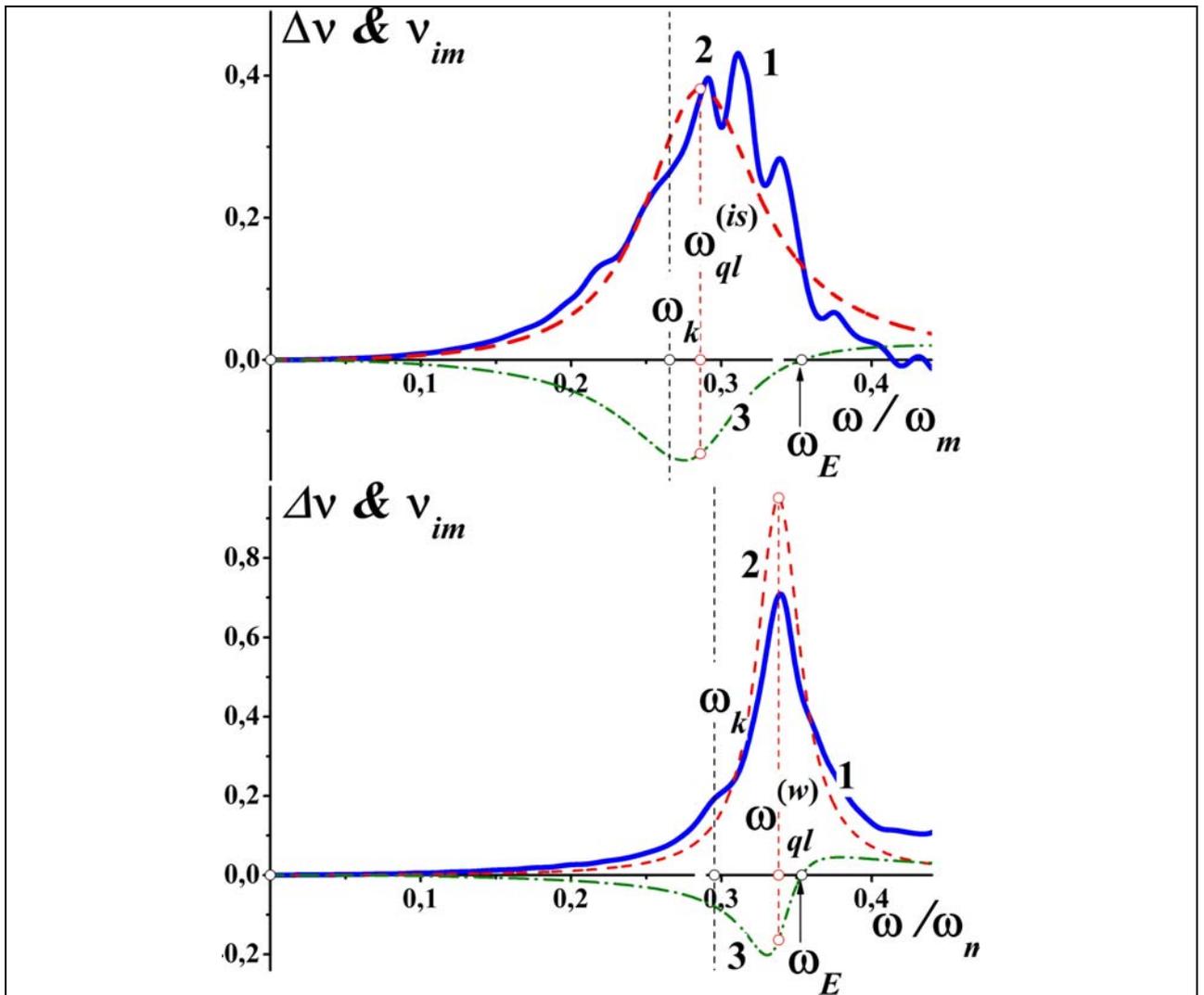

**Fig. 3** Phonon density for the impurity concentration of 5% (curves 1), spectral densities of isolated impurity atoms (curves 2); spectral correlators of impurity with its nearest environment (curves 3). the upper and bottom panels correspond to the heavy isotopic and weakly coupled impurities, respectively

Fig. 4 shows the evolution of the phonon densities of states of disordered solid solutions with increasing concentration of heavy isotopic impurity. We present the phonon densities of states $\nu(\omega)$ for concentrations $p = 0.05$; $0.10$; $0.25$ and $0.5$ (solid lines in the relevant fragments of the picture). Along with these curves the dependencies $\nu(\omega, p)$ of the perfect crystal-matrix and also of an ideal lattice, consisting of heavy ($\varepsilon = 3$) atoms defined as the isotopic defect when describing the solution (thin dashed lines), are shown in each fragment as some reference point. The spectral density of the system $\rho(\omega)$ is a self-averaging quantity (see, e.g., [17]) and can be obtained by averaging over all provisions $r$ and directions of displacement $i$ of function $\rho_i(\omega, r)$. $\rho_i(\omega, r)$ are the spectral densities in the cyclic subspaces generated by the displacements of atoms with the radius vector $r$ in the crystallographic direction $i$. We calculated spectral densities $\nu(\omega, p) \equiv \langle \rho_i(\omega, r) \rangle_p$ for different concentrations of randomly distributed impurity atoms. For each concentration value averaging was carried out over several thousand random configurations of the impurity distribution. For each configuration the density of states was determined by averaging of several tens of spectral densities corresponding to displacements along different crystallographic directions of a few tens of consecutive atoms.

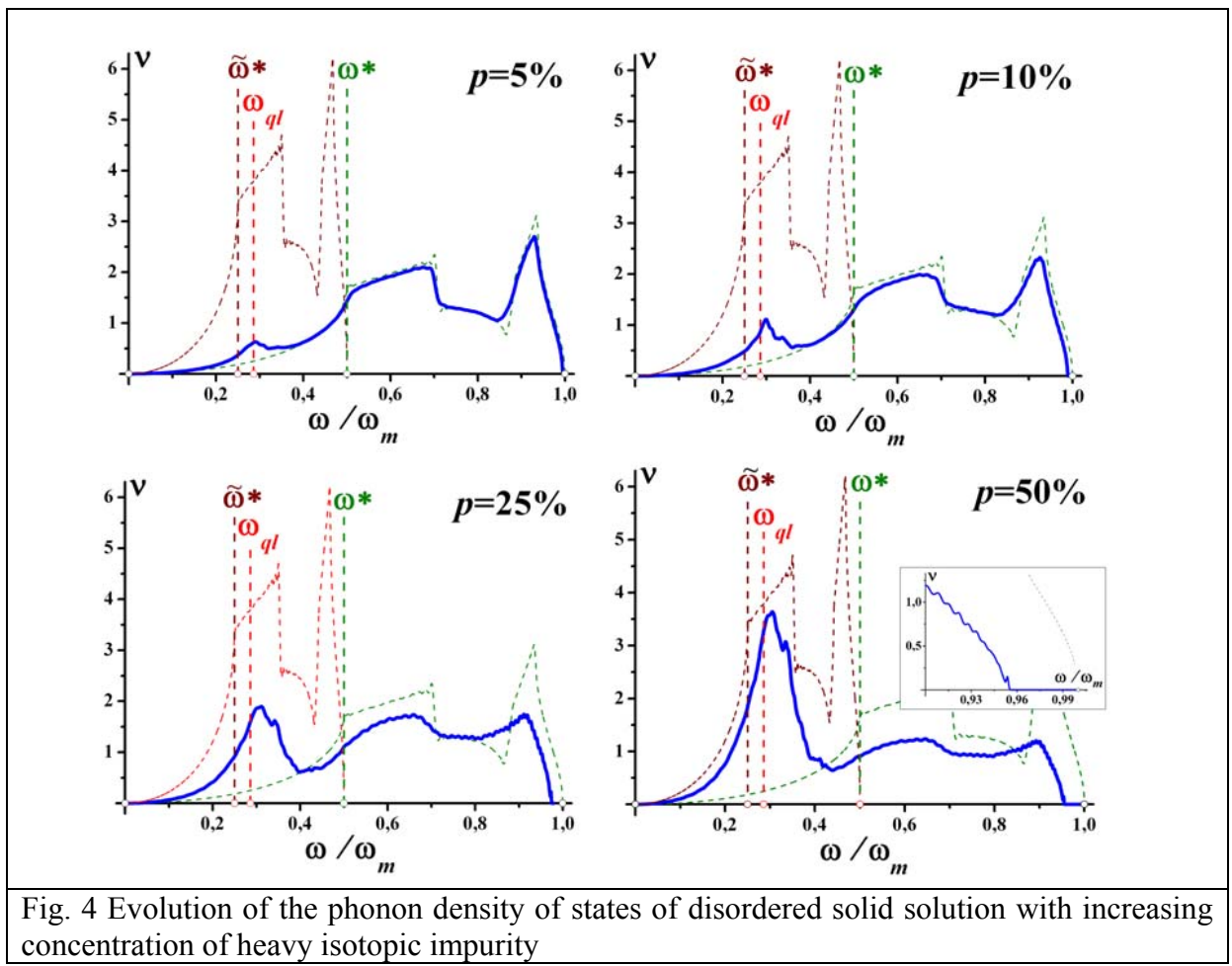

Fig. 4 Evolution of the phonon density of states of disordered solid solution with increasing concentration of heavy isotopic impurity

It should be noted here that even for $p = 0.05$ the crystal regularity of the atoms arrangement is substantially impaired and it is impossible to use the term "*the first Brillouin zone*" for the description of the quasi-particle spectra of such a solution. At the same time the function $\nu(\omega, p)$ near the frequency $\omega^*$ qualitatively changes its behaviour not only when $p = 0.05$ but also at $p = 0.10$, and this change is completely analogous to change of the behaviour of phonon density of states near the first van Hove feature. Let us recall that the first van Hove singularity corresponds to the change in the topology of isofrequency surfaces i.e. to the transition from closed surfaces (when $\omega < \omega^*$) to open along a certain direction in reciprocal space ones (at $\omega > \omega^*$) (see, e.g., [1,2]) . This transition is due to touch of isofrequency to surface boundary of the first Brillouin zone, and in the absence of the crystal translational symmetry it is impossible to talk about it.

In the next subsection a connection between the group velocities of acoustic phonon modes and van Hove features will be analyzed in more detail. Here we note that localization of phonons with certain polarization along some directions which occurs at specific frequencies is not directly related to the translational symmetry of the crystal, but will be valid also for disordered systems. For ideal crystal the frequencies of the "braking" of next phonon mode are the same for the entire sample and exhibit themselves in the behaviour of the phonon density as singularities whose form is determined by the dimension of the lattice. For structure with broken translational symmetry every "braking" mentioned earlier occurs within a certain frequency range and the correspondent singularities will be smoothed out.

Even at $p = 0.10$ the quasi-local peak has a certain structure that remotely resembles the phonon density of states of FCC crystal, determined using a small number of moments i.e. the J-matrix of rank $3 \div 5$ [48]. This is due to the formation of impurity clusters of appropriate size. The left slope of the peak gets parabolic shape and quasi-local frequency becomes the new boundary of propagon zone. At the same time on the phonon density it is still possible to identify high-frequency van Hove features of host lattice. With increasing concentration amount of impurity clusters increases and the boundary of propagon zone becomes more distinct. When $p = 0.25$ this boundary is already looks as a fracture, similar to the first van Hove feature in the regular crystal. When $p = 0.5$ the transformation of quasi-local maximum into the spectrum of the lattice of heavy atoms is substantially completed. One can clearly see two van Hove singularities: first one, separating propagons from diffuzons and another one above which phonons are propagating very slowly (locon region). In this area the curve $\nu(\omega, 0.5)$ looks like fractal curve (similar to that obtained in [53]) for a one-dimensional solid solution). It is clearly seen in Fig. 4 (see inset on the fragment $p = 0.5$) that the spectrum ends by the exponential damping of the vibrations, characteristic for disordered systems [17]. The fact that the maximum frequency in spectra shown in Fig. 4 differs from the so-called natural boundary (in this case - the maximum frequency of the host lattice) can be explained by the fact that the rank of calculated J-matrixes is finite ($n = 76$) and does not allow an arbitrarily large cluster size of both the atoms of the host lattice and impurity.

Thus the condition for the existence of QLV in propagon zone is the existence of solutions of equation (3) in this frequency range. The frequency of quasi-local vibration is determined by the frequency of the maximum on the imaginary part of the local Green function of the impurity atom. Propagation of quasi-local vibrations occurs with very low velocities and can be presented as diverging waves. Scattering of quasi-plane acoustic waves on mentioned diverging waves occurs at finite impurity concentrations $p$ and starting from $p \sim 10\%$ quasi-local frequency becomes the boundary of propagon zone of the phonon spectrum of solution.

## 3. The interaction of acoustic phonons with quasi-localized oscillations. Van Hove singularities and boson peaks

Let us analyze in more detail connection of Van Hove singularities in the phonon spectrum of the ideal crystal and similar features of the phonon spectra of structures with broken crystal regularity in the arrangement of atoms with a dispersion of phonons. For any sold (both crystal and the one which does not possess the translational symmetry of the atoms arrangement), a low-frequency range exists where the dispersion low of phonons $\omega(k) = s(\kappa)k$ will be acoustic ($\kappa \equiv \frac{k}{k}$, and $s(\kappa)$ is the velocity of sound). Phonon density of states in this range has the Debye form $\nu(\omega) \sim \omega^2$.

With increase of $k$ the phonon dispersion law increasingly deviates from the linear (frequency $\omega$ becomes to be less than $sk$) and the actual density of states deviates from the Debye one upwards.

Boson peaks can be regarded as maxima on the ratio $\frac{\nu(\omega)}{\omega^2}$ only when $\omega < \omega^*$ because the maximum on the mentioned ratio correspondent to the first van Hove singularity ($\omega = \omega^*$) always exists. In given frequency range (propagon zone) phonon density can be approximated by a parabola, and its deviation from the Debye density $\nu_D(\omega)$ can be expressed in terms of frequency dependence of the value $\omega_D$ i.e. the phonon density could be written as follows

$$\nu(\omega) \equiv \frac{3\omega^2}{\omega_D^3(\omega)} \qquad (10)$$

Then, using the definition of $\omega_D$ (see, e.g., [54]), the ratio of the phonon density to the squared frequency can be expressed through the dispersions of sound velocities $s_i(\omega)$

$$\frac{\nu(\omega)}{\omega^2} \equiv \frac{3}{\omega_D^3(\omega)} = \frac{V_0}{6\pi^2} \sum_{i=1}^{3} \frac{1}{s_i^3(\omega)} \qquad (11)$$

($V_0$ is unit cell volume). Thus, the occurrence of the maximum on the ratio $\frac{\nu(\omega)}{\omega^2}$ is due to the additional dispersion of sound velocities. This dispersion could be caused by the complicated structure of the unit cell and structure heterogeneities, which are the source of quasi-localized vibrations (defects, rotational degrees of freedom of lattice, etc.).

In a perfect crystal, where the dependence $\omega(k)$ is periodic function, at a certain $k = k^*$ group velocity of phonons of some transverse acoustic modes for one of the crystallographic directions vanishes. Typically, this direction coincides with the direction of one of the axes of symmetry in $k$-space, and the value $k^*$ corresponds to the boundary of the first Brillouin zone in this direction. I.e. a transition from closed to open isofrequency surfaces occurs at the frequency $\omega^* = \omega(k^*)$, and the value $\omega^*$ is the frequency of the first Van Hove singularity.

Fig. 5 shows the density of states of the FCC lattice with central nearest-neighbour interaction and dependencies on the frequency of the group velocities of longitudinal and transverse phonons along high-symmetry crystallographic directions ΓX, ΓL and ΓK ($s_0 = \frac{a\omega_m}{4}$). The first octant of the first Brillouin zone of FCC crystal is shown on the right for convenience. It could be seen that the Debye density of states $\nu_D(\omega) = 3\omega^2 \omega_D^{-3}$, depicted in Fig. 5 by thin dashed line, quite satisfactorily coincides with the true density of states at $\omega \leq 0.25\omega_m$.

In the case $\omega > 0.25\omega_m$ the curve $\nu(\omega)$ starts to deviate from the curve $\nu_D(\omega)$ upwards. The van Hove singularity corresponds to the vanishing of the group velocity of phonons of transverse modes propagating along the direction ΓL. For this model the frequency $\omega^* = \frac{\omega_m}{2}$. In the case $\omega \geq \omega^*$ the nature of the phonon dispersion varies qualitatively. Together with rapidly propagating phonons whose dispersion law is still close to linear, localized states appear so as along ΓL direction the transverse phonons no longer propagate. As could be seen from Fig. 5, dependence $s(\omega)$ for the remaining branches and propagation directions at $\omega \approx \omega^*$ is relatively low. Average group velocity of phonons for $\omega < \omega^*$ gradually decreases with increasing frequency, and at $\omega = \omega^*$ its further decrease occurs abruptly.

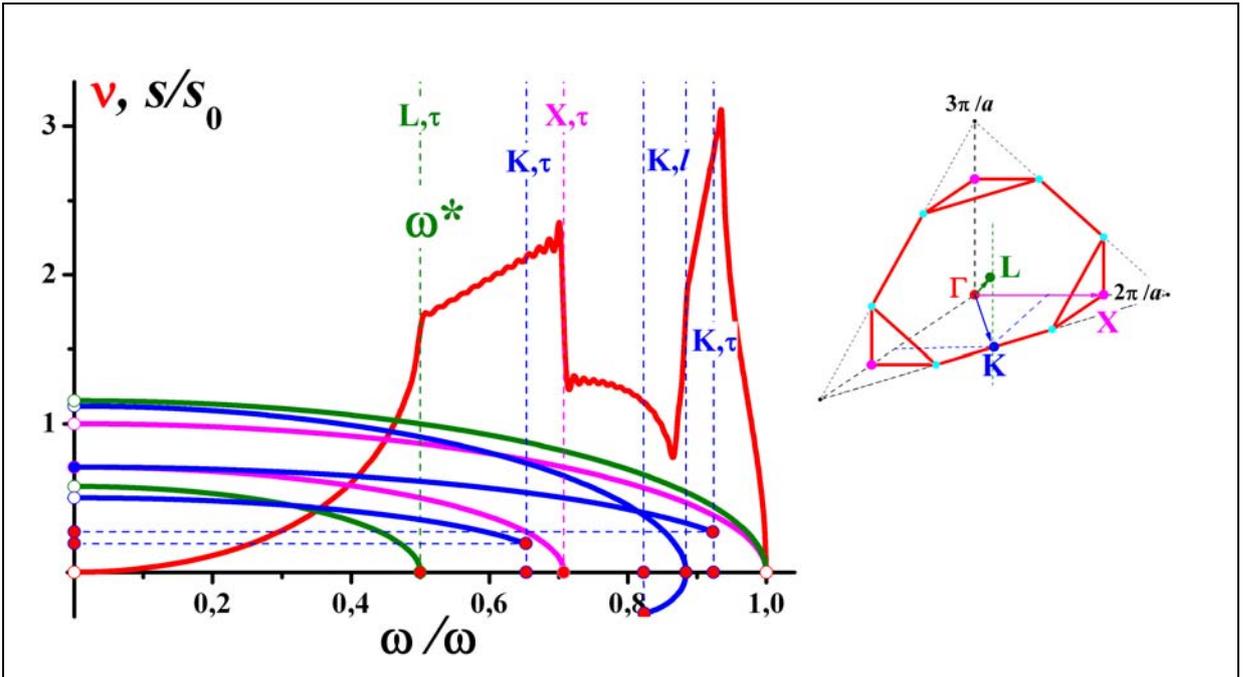

**Fig. 5** The phonon density of states (thick solid line) and the frequency dependencies of the group velocities of phonon modes (thin lines) along the main of the highly symmetrical crystallographic directions of FCC crystal with central nearest-neighbour interaction. Indices $l$ and $\tau$ corresponds to longitudinal and transverse modes respectively.

Presentation of the phonons dispersion low as a dependence of their velocity $s_i(\omega)$ on frequency can be more naturally generalized to disordered systems than dependence $\omega(k)$. The frequency of the first the van Hove feature is the interface between fast and slow phonons (propagons and diffuzons) in perfectly ordered crystal. In terms of [30] the mentioned frequency can be considered as an analogue of the Ioffe-Regel crossover in a regular crystal system: when $\omega = \omega^*$ the average wavelength exceeds the interatomic distance. Let us examine the effective wavelength of the phonon having frequency $\omega$, i.e., the distance covered by phonon over its vibration period $\lambda_{eff}(\omega) = \dfrac{2\pi|s(\omega)|}{\omega}$. Frequency dependencies $\lambda_{eff}(\omega)$ are presented in Fig. 6.

It could be seen that even when the $p=1\%$, value $\lambda_{eff}(\omega_{ql})$ exceeds the average distance between impurity atoms and even at these concentrations of the impurity scattering of acoustic phonons on quasi-localized vibrations will have a Ioffe-Regel nature and manifest itself in the behaviour of spectral densities in a form similar to the one of the first van Hove singularity in ideal crystal. Let us note also that at $\omega^{**}$ (the frequency of high-frequency van Hove feature) correspondent to the transition from open to closed isofrequency surfaces, the value $\lambda_{eff}(\omega)$ for the most long-wave vibrations (longitudinal vibrations propagating along the direction $\Gamma L$) becomes equal to the distance between the nearest neighbours $l_0 = \dfrac{a}{\sqrt{2}}$. I.e. in the case when $\omega > \omega^{**}$ the phonons can be regarded as quasi-localized states. This frequency range is called "*local zone*", and phonons with frequencies exceeding $\omega^{**}$ are called "*lockons*" (see, e.g., [55]).

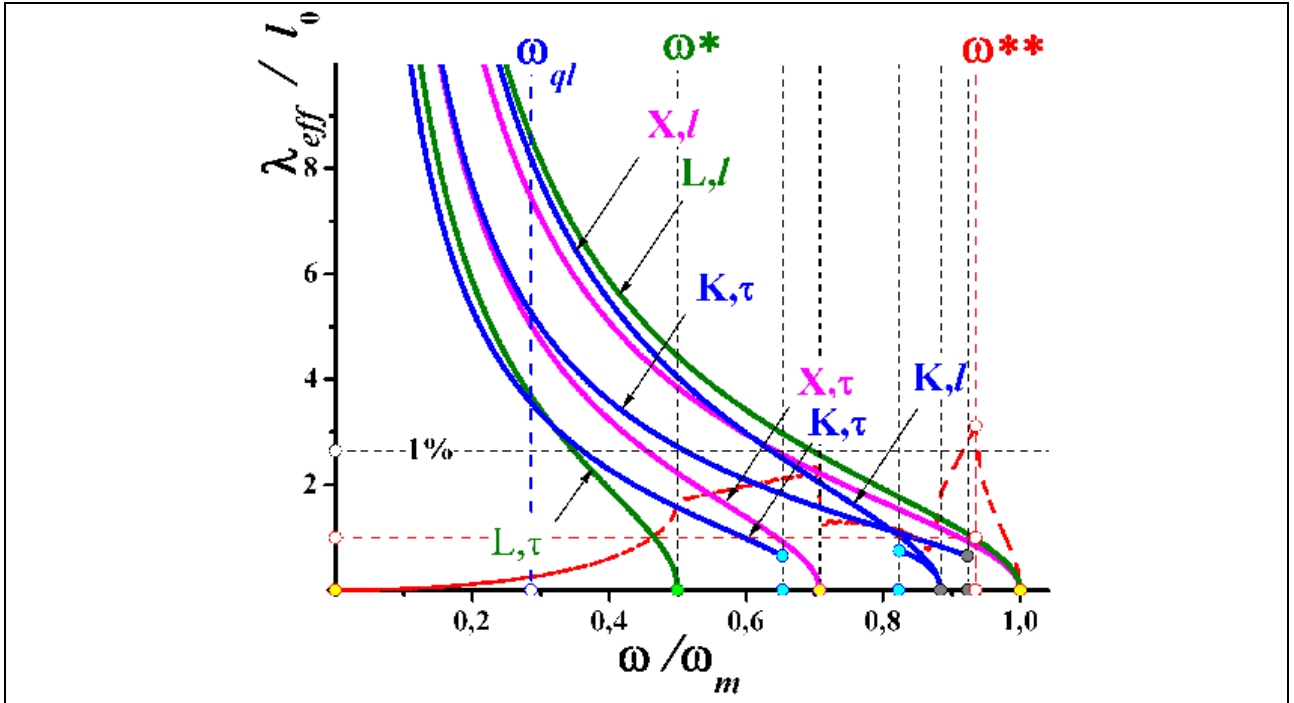

**Fig. 6** The phonon density of states (dashed line) and frequency dependencies of the values $\lambda_{eff}(\omega)/l_0$ (heavy solid lines) along the main highly symmetric crystallographic directions of FCC crystal with central nearest-neighbour interaction. Designations are the same as in Fig. 5

Let us investigate the manifestations of the boson peaks and the Ioffe-Regel crossover in phonon spectra of solid solutions. Fig. 7 shows the contributions of impurity atoms (curves 4) and the atoms of lattice-matrix (curves 5) to the phonon densities of states of 5% solid solutions (curves 3) of the heavy isotope (left fragment) and weakly bound impurities (right part).

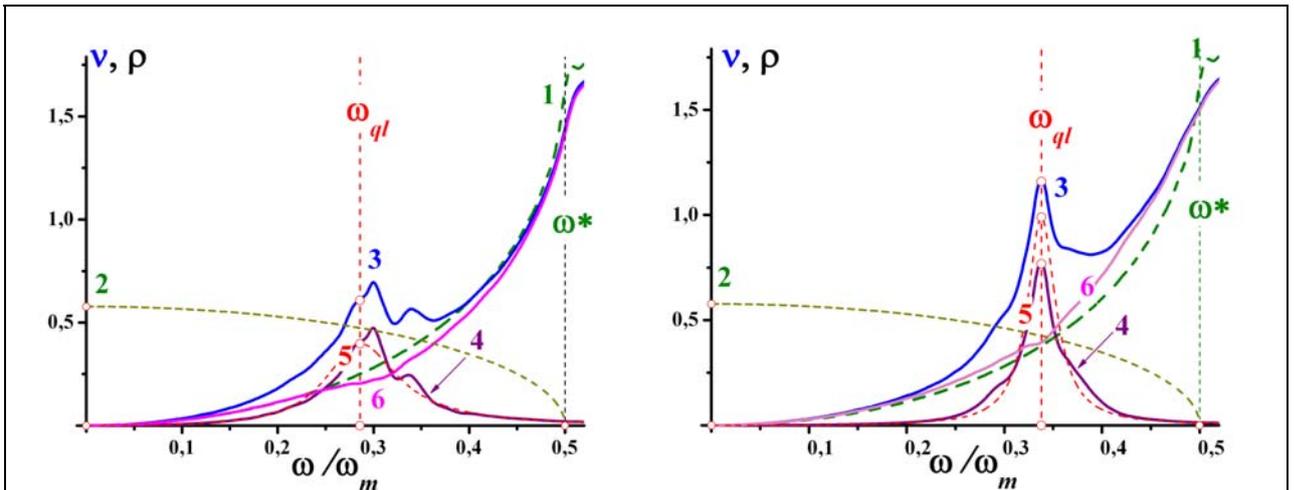

Fig. 7 Contributions of impurity atoms and host lattice atoms in the phonon densities of states of solid solutions.

Curves 1 represent phonon density of states of the ideal crystal, curves 2 are the frequency dependencies of the transverse sound velocity in the direction $\Gamma L$; thin dashed curves are the spectral densities $\rho^{(\tau_-^5)}(\omega)$.

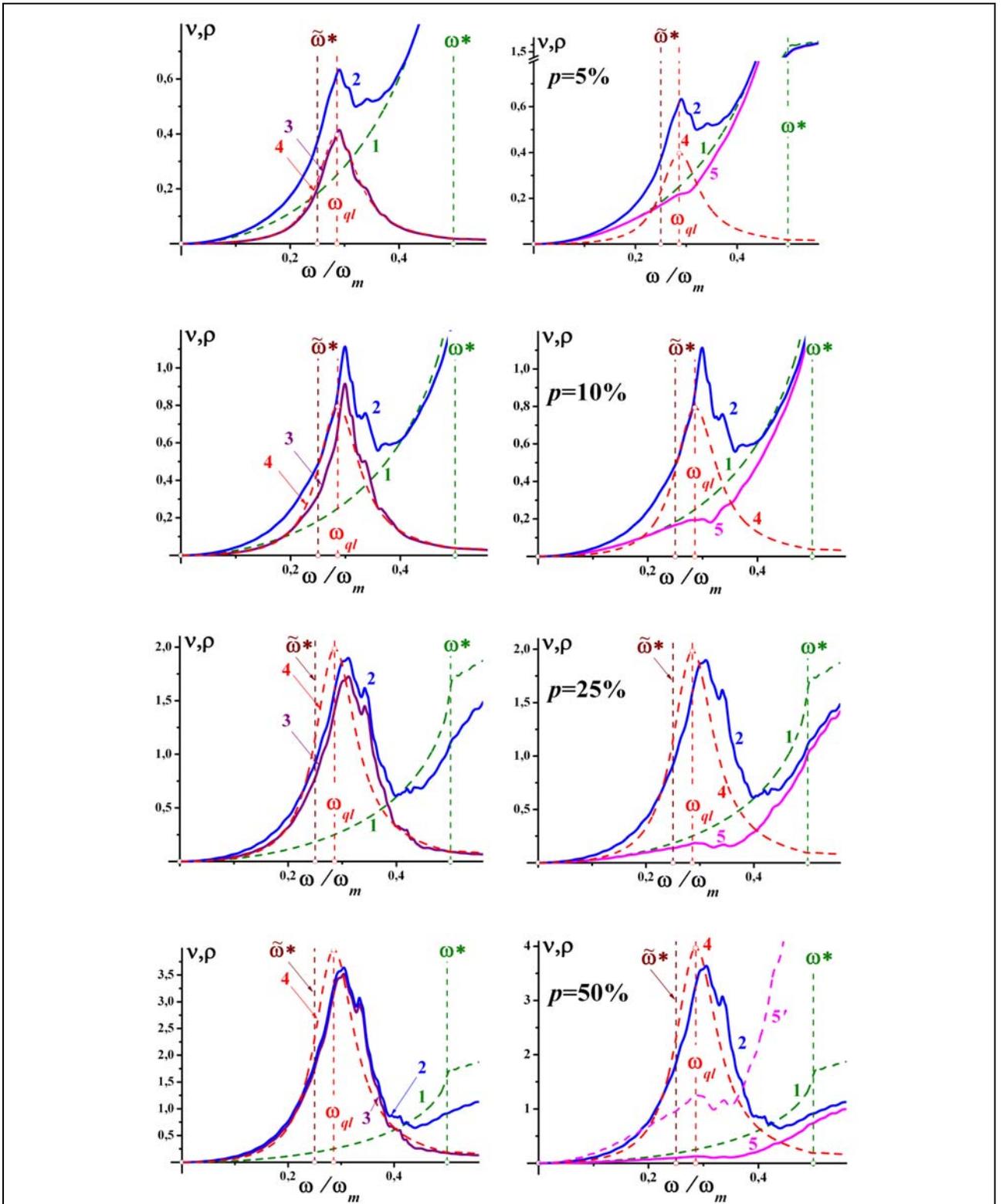

Fig. 8 The evolution of deposits from the vibrations of impurity atoms (left) and from the atoms of the host lattice (right) in the phonon density of states of solid solution with increasing of impurity concentration

One can see that quasi-local maxima formed on the phonon density $\tilde{\nu}(\omega)$ (curves 3) entirely caused by vibrations of impurity atoms, whose contribution $\nu_{imp}(\omega)$ represented in the Fig.7 as curves 4. Function $\nu_{imp}(\omega)$ differs from zero only near the maximum of the curve $\rho^{(\tau_-^5)}(\omega)$ (frequency $\omega_{ql}$). Therefore, QLV can be presented as slowly diverging from the impurity wave similar to spherical.

Vibrations of the atoms of the host lattice in the given frequency range propagate rapidly and become scattered by the QLV. Curves 5 in Fig. 7 are depicting the frequency dependence $\tilde{\nu}(\omega) - \nu_{imp}(\omega)$. At frequencies $\omega < \omega_{ql}$ vibrations mentioned earlier propagate as plane waves and the corresponding areas on the curves 5 are smooth and have parabolic (quasi Debay) form. For $\omega \approx \omega_{ql}$ a kink similar to the shape of the first van Hove feature can be seen on these curves. At this frequency, as well as for $\omega = \omega^*$ an abrupt change of the average group velocity of phonons occurs in the phonon spectrum of an ideal crystal (Fig. 5). The wave-length equal to the average distance between impurities corresponds to the frequency $\omega_{ql}$ and it is larger than the one corresponding to the frequency $\omega^*$. The frequency $\omega_{ql}$ is the upper limit of propagon zone of solid solution, which is clearly exhibiting with increasing of its concentration.

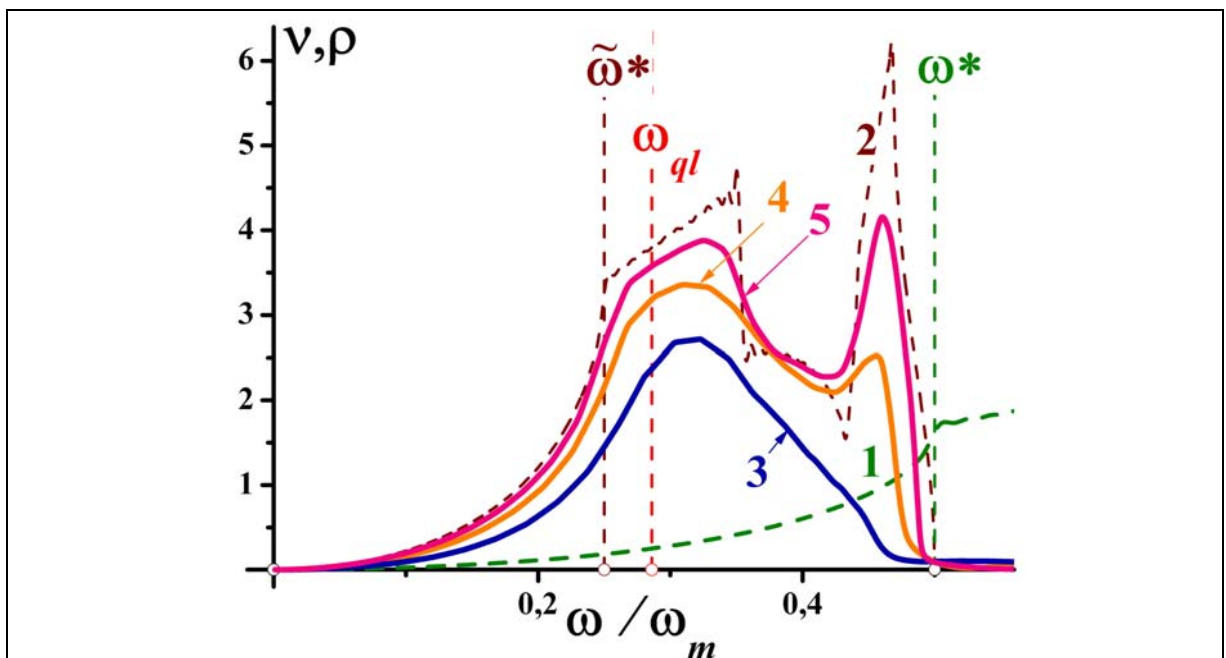

Fig. 9 Low-frequency regions of contributions to the phonon density of states of solid solutions with a high impurity concentration of impurity atoms. Curve 1 (dashed line) shows phonon density of states of the ideal lattice, curve 2 (dashed line) shows it for ideal lattice consisting of heavy atoms, curve 3 (solid) is for p = 50%, curve 4 (solid) is for p = 75%; curve 5 (solid) is for p = 90%.

Fig. 8 shows the evolution of deposits in the phonon density of states from the displacements of impurity atoms (curves 3 in the left fragments of the picture) and from the displacements of atoms of the host lattice (curves 5 in the right part) with increasing impurity concentration. For all fragments: the curves 1 (dashed) are phonon densities of states of the ideal host lattice; curves 2 are phonon densities of states of the solution; curves 4 are values $p\tilde{\rho}^{(\tau_-^5)}(\omega)$, where $\tilde{\rho}^{(\tau_-^5)}(\omega)$ is the spectral density in the invariant subspace generated by the displacement of an isolated impurity atom (8). It is clear that even at high impurity concentrations curves 5 have a characteristic kink similar to the shape of the first the van Hove feature at $\omega \approx \omega_{ql}$. This kink corresponds to the transition from fast-propagating phonons (propagons) to slower ones (diffuzons). Thus, both kinks in the curves 5 (Fig. 7 and Fig. 8) and the first van Hove singularity have a common nature: they are caused by an abrupt change of the average group velocity of phonons and are manifestations of the Ioffe-Regel crossovers. In curves 3, such a transition from propagons to diffuzons occurs near the quasi-local frequency $\omega_{ql}$ (see also Fig. 4).

Note that the frequency of this transition $\omega \geq \omega_{ql}$, while the frequency of the first van Hove singularity of the crystal at $p = 100\%$ is $\tilde{\omega}^* < \omega_{ql}$. Together with further increase in concentration impurity clusters of sufficiently large size are formed in the solid solution. Various crystallographic directions can be identified in these clusters. Frequency $\tilde{\omega}^*$ corresponds to the vanishing group velocity of the transversely polarized phonons in mentioned clusters along the crystallographic direction $\Gamma L$. The evolution of low-frequency parts of the contributions to the phonon density of states from impurity atoms is shown in Fig. 9. Concentration $p$ changes from 50% to 100%.

Thus the influence of impurity atoms, which are heavy or weakly bound to the atoms of host lattice, on the phonon spectrum and the vibrational characteristics is manifested both in the formation of quasilocal vibrations caused by the vibrations of impurities, and in the scattering on these vibrations of fast acoustic phonons generated by atomic vibrations of the host lattice.

## 4. Conclusions

The present work shows the common nature of van Hove singularities with the phonon Ioffe-Regel crossovers and Boson peaks. The statement of the experimental proof of the similar nature of both BP appearance in glasses and van Hove singularities in crystals was made in [56-58]. It was established that these anomalies of the phonon spectrum are due to the additional group-velocity dispersion of rapidly propagating phonons (the so-called propagons) on slow quasi-particles. Formation of the boson peaks during the scattering of acoustic phonons on quasilocalized vibrations in disordered solid solutions was analyzed at the microscopic level. It was shown that in the frequency range below the frequency of the first van Hove singularity (propagon zone) the singularities of kink type (i.e. the singularities looking similarly to the mentioned feature) will be formed on the phonon density of states.

Enrichment of low-frequency part of phonon spectrum is due to not only the formation of quasi-localized states, but also to decrease of the propagation velocity of long-wavelength acoustic phonons happening due to their scattering by these states. To ensure that the results of this slowdown is clearly manifested as the boson peaks on the ratio $\frac{\nu(\omega)}{\omega^2}$ or as additional singularities of Ioffe-Regel crossovers type in the frequency range $[0, \omega^*]$, it is necessary to satisfy the following conditions. First, the frequencies of scattering of quasi-local vibrations have to be sufficiently low, i.e. "the power of the defect" should be sufficiently large. Secondly, the size of the defect cluster should be large enough (at least two interatomic distances) which requires a sufficiently high (15-20%) concentration of defects. The second condition means the appearance of one more parameter with dimension of length $l \sim \frac{2\pi s_0}{\omega} > a$ ($a$ is interatomic distance) in the system.

Parameter $l$ plays in our case the role of the parameter of disorder. In this case, as noted in [28], the continuum approximation is not applicable even for description of the long-wavelength phonons.

This work was partially supported by grant # 23/09-N NAS